\documentclass[aps,pra,10pt,twocolumn,showpacs, superscriptaddress]{revtex4-1} 
\usepackage{amsfonts,amsmath,amssymb}
\usepackage{graphicx}
\usepackage[pdfstartview=FitH,colorlinks=true,linkcolor=blue,citecolor=blue,urlcolor=blue]{hyperref}
\usepackage{tikz}
\usepackage{verbatim}
\usepackage{braket}

\begin{document}

\title{Hydrodynamic VS collisionless dynamics of a 1D harmonically trapped Bose gas}

\author{Giulia De Rosi}
\email{giulia.derosi@unitn.it}
\affiliation{INO-CNR BEC Center and Dipartimento di Fisica, Universit\`a di Trento, Via Sommarive 14, I-38123 Povo, Italy}


\author{Sandro Stringari}
\email{stringar@science.unitn.it}
\affiliation{INO-CNR BEC Center and Dipartimento di Fisica, Universit\`a di Trento, Via Sommarive 14, I-38123 Povo, Italy}

\date{\today}

\begin{abstract}
By using a sum rule approach we investigate the transition between the hydrodynamic and the collisionless regime of the collective modes in a 1D harmonically trapped Bose gas. Both the weakly interacting gas and  the  Tonks-Girardeau limits are considered.  We predict that the excitation of the dipole compression mode is characterized, in the high temperature collisionless regime, by a beating signal of two different frequencies ($\omega_z$ and $3\omega_z$) while, in the high temperature collisional  regime,  the excitation consists of a single frequency ($\sqrt{7}\omega_z$). This behaviour differs from the case of the lowest breathing mode whose excitation consists of a single frequency ($2\omega_z$) in both regimes. Our predictions for the dipole compression mode open promising perspectives for the experimental investigation of  collisional effects in 1D configurations. 
\end{abstract}

\pacs{PACS numbers}

\maketitle

\section{Introduction}

Thermalization and relaxation phenomena  represent a key issue in one-dimensional (1D) systems \cite{Popov2001, Giamarchi2004} of identical bosons with zero--range repulsive interaction due to the intrinsic integrability \cite{Thacker1981, Yurovsky2008, Rigol2007} of this many-body system and have been the object of recent experimental and theoretical investigations \cite{Kinoshita2006, Hofferberth2007, Hofferberth2008, vanAmerongen2008, Mazets2008, Mazets2009, Tan2010, Laburthe Tolra2004}. They  play an important role not only for the achievement of equilibrium but also for the propagation of collective modes \cite{Mazets2011} whose nature, in harmonically trapped configurations, is expected to evolve from the hydrodynamic regime (HD) at low temperature to a collisionless (CL) regime at higher temperature. At low temperature, the applicability of the hydrodynamic description is ensured by the phononic nature of the elementary excitations. Phonons are in fact known to characterize the long wavelength dispersion of  the excitation spectrum in one-dimensional interacting Bose gases \cite{Lieb1963} and their description has the same form as the one given by the hydrodynamic theory of superfluids \cite{Menotti2002, DeRosi2015}.
At high temperature \cite{highT}, due to the exponential decrease of the density caused by harmonic trapping, collisions become rare and the system enters the collisionless regime described by  the ideal gas model. One then expects a transition between the two regimes which could provide valuable informations on the collisional effects in 1D configurations.

So far most of the attention in the collective features of 1D harmonically trapped Bose gases  has concerned  the lowest breathing (LB) mode. The frequency of this mode was calculated at T=0 within the Lieb-Liniger model using a sum rule approach \cite{Menotti2002}, exploring the transition from  the weakly interacting Bogoliubov gas  (BG) to the Tonks-Girardeau (TG) limit of strongly repulsive bosons \cite{Tonks1936, Girardeau1960}. The experimental results of \cite{Haller2009} have confirmed with good accuracy the predictions of theory. Recent studies  of this mode have also focused on the so called super-Tonks-Girardeau (STG) regime of hard rods \cite{Astrakharchik2005, Haller2009} and on the regime of small number of particles (or small coupling constant $g_{1D}$) where the Local Density Approximation (LDA), usually employed to calculate the density profiles using the equation of state of uniform matter, is not applicable \cite{Gudyma2015, Chen2015, Gudyma20152}. The temperature dependence of the frequency of the lowest breathing  mode has also been the object of recent  theoretical \cite{Hu2014, Chen2015} and experimental \cite{Moritz2003, Fang2014} work. The theoretical predictions are usually based on a hydrodynamic description where the relevant thermodynamic quantities are calculated using the Yang-Yang theory \cite{Yang1969}, which generalizes the Lieb-Liniger theory \cite{Lieb1963} of interacting 1D bosons to finite temperature. A characteristic feature of the hydrodynamic theory applied to the lowest breathing mode is that, at high temperatures,  it predicts \cite{Fang2014, Hu2014, DeRosi2015} the same frequency $\omega= 2\omega_z$ as given by the non interacting gas model, see Table \ref{Tab:monopoledeep}. This rules out the possibility of a simple identification of the hydrodynamic VS the collisionless nature of the oscillation. 

In this work we exploit the different behaviour exhibited by the dipole compression (DC) mode, identified as the  lowest compression mode with the same parity as the center of mass (dipole) mode. Differently from the center of mass mode,  which oscillates with the model independent  frequency $\omega=\omega_z$,  the dipole compression mode is sensitive to the equation of state and, differently from the lowest breathing  mode,  is characterized by a different excitation spectrum at high temperatures, when investigated in the hydrodynamic or in the collisionless regimes, see Table \ref{Tab:monopole2}. This mode, whose frequency has been already measured at low temperature in elongated configurations in the case of the unitary Fermi gas \cite{Tey2013},  is consequently  a natural candidate to exploit the effects of relaxation caused by collisions and the corresponding thermalization effects in 1D configurations. 
Numerical calculations for the DC frequencies at zero and finite temperature in the hydrodynamic framework have been carried out in \cite{Hu2014}.

In the following we will use  the Lieb-Liniger Hamiltonian  \cite{Kheruntsyan2005}
\begin{multline}
\label{Eq:H}
H = H_{kin} + H_{int} + H_{trap} = \\ = -\frac{\hbar^2}{2m} \sum_{i = 1}^N \frac{\partial^2}{\partial z_i^2} + g_{1D}\sum_{i > j}^N\delta(z_{ij}) + \sum_{i = 1}^NV_{ext}(z_i)\ ,
\end{multline} 
describing a gas of 1D interacting Bose particles in the presence of the harmonic potential $V_{ext}(z) = m\omega_z^2 z^2/2$. Here $z_{ij}\equiv z_i - z_j$ is the relative coordinate and $g_{1D}$ is the relevant 1D coupling constant.  In the presence of radial harmonic trapping and in the absence of confinement induced resonance  \cite{Olshanii1998,Petrov2000}, the interaction parameter $g_{1D}$ can be written as  $g_{1D}=2\hbar^2a/ma^2_{\perp}$  where $a$ is the three-dimensional scattering length and $a_\perp$ is the radial oscillator length.

Our paper is organized as follows.

In Section \ref{Sec:variational} we summarize the basic results of hydrodynamic theory of 1D gases confined by a harmonic potential. This theory allows for analytic results for the collective frequencies  if the equation of state exhibits a polytropic dependence on the density \cite{DeRosi2015}. Furthermore it can be conveniently formulated using a variational procedure allowing for an easy determination of the collective frequencies in the intermediate regimes of temperature and interaction. 
 
In Section \ref{Sec:sumrules} we formulate  a sum rule approach to describe the frequency of the collective oscillations in the presence of harmonic trapping. This approach provides a useful insight on the physical features of  the collective oscillations, both at zero and finite temperature. In this Section we will  also provide a valuable derivation of  the 1D virial theorem, holding in all  regimes of temperature and interaction. An extension of the virial theorem, which turns out to be useful for the study of the dipole compression mode, will be also presented.

In Section \ref{Sec:Compressiondipole} we discuss the dipole compression frequency and  point out the different behaviour exhibited in the hydrodynamic and in the collisionless regime of high temperature. In particular, in the latter case, this mode exhibits a characteristic beating effect involving two different frequencies which are expected to be of easy experimental identification. 

In Section \ref{Sec:conclusion} we draw our final conclusions.

\section{Hydrodynamic theory of 1D Bose gases in the presence of harmonic trapping} 
\label{Sec:variational}

We consider the 1D version 
\begin{equation}
\label{Eq:wave equation}
m(\omega^2-\omega^2_z)n v + \frac{\partial}{\partial z} \left[n\left(\frac{\partial P}{\partial n}\right)_{\bar{s}} \frac{\partial v}{\partial z}\right] =0
\end{equation}
of the linearized hydrodynamic equation \cite{Griffin1997, Taylor2009, DeRosi2015}
 for the velocity field $v(z)$, 
where $(\partial P/\partial n)_{\bar{s}}$ is the adiabatic compressibility ($\bar{s}$ being the entropy per particle) evaluated at the local value of the 1D equilibrium density profile $n \equiv n(z)$ 
 whose $z$-dependence, caused by the external potentials $V_{ext}(z)$, can be determined in the Local Density Approximation, through the    solution of the equilibrium  Euler equation
\begin{equation}
\label{Eq:Euler}
\left(\frac{\partial P(z)}{\partial n}\right)_T\frac{\partial n(z)}{\partial z} +n(z)\frac{\partial V_{ext}(z)}{\partial z}  =0 \ ,
\end{equation}
 for a fixed value of the temperature of the gas.

The above equations show that  the eigenfrequencies $\omega$ of the collective oscillations  are determined once the adiabatic and the isothermal $(\partial P/\partial n)_{T}$ compressibilities, calculated at the local value $n(z)$ of the density, are known. These quantities depend on the interaction and on the temperature of the gas.

In the uniform case ($V_{ext}=0$) Eq. \eqref{Eq:wave equation} admits a plane wave solution  $v \propto e^{iqz}$ yielding the  phonon dispersion relation $\omega=c_sq$, where $c_s=\sqrt{(\partial P/\partial n)_{\bar{s}}/m}$ is the  adiabatic  sound velocity.

It is worth noticing that, since in 1D there is no superfluid phase transition  \cite{Mermin1966, Hohenberg1967},  Eq. \eqref{Eq:wave equation} can be applied to all temperatures provided the dynamic behaviour of the gas is correctly  described by hydrodynamic theory. This represents an important difference with respect to 2D and 3D systems  where hydrodynamic theory, for temperatures below the critical value, should be generalized to the Landau theory of two fluids \cite{Pitaevskii2016}. 

It is immediate to show that Eq. \eqref{Eq:wave equation} can be derived \cite{Hou2013} from the variational approach $\delta \omega^2/ \delta v=0$ , with 
\begin{equation}
\label{Eq:variational}
\omega^2 = \omega_z^2 + \frac{\int dz n \left(\frac{\partial P}{\partial n}\right)_{\bar{s}} \left(\frac{\partial v}{\partial z}\right)^2} {\int dz m n v^2}\ ,
\end{equation}
first developed in 3D systems \cite{Taylor2005, Taylor2008, Taylor2009}. 
The advantage of using the variational approach, Eq. \eqref{Eq:variational}, rather than the differential hydrodynamic equation, Eq. \eqref{Eq:wave equation}, is that one can easily estimate the collective frequencies, at zero as well as at finite temperature, with a suitable ansatz for  the velocity field. This method has been recently  implemented in \cite{Hu2014}.

In addition to the universal dipole result $\omega(D)=\omega_z$ for the center of mass oscillation (Kohn mode), corresponding to the choice $v= const$, useful expressions for the frequencies of the  relevant collective modes concern the lowest breathing mode
\begin{equation}
\label{Eq:variationalLB}
\omega^2_{HD}(LB) = \omega_z^2 + \frac{\int dz n \left(\frac{\partial P}{\partial n}\right)_{\bar{s}}} {\int dz m n z^2}\ ,
\end{equation}
corresponding to the ansatz $v =z$, and
the dipole compression mode 
\begin{equation}
\label{Eq:variationalDC}
\omega^2_{HD}(DC) = \omega_z^2 + \frac{\int dz n \left(\frac{\partial P}{\partial n}\right)_{\bar{s}}4z^2} {\int dz m n (z^2-\langle z^2\rangle)^2}\ ,
\end{equation}
corresponding to the ansatz $v =z^2-\langle z^2\rangle$ where $\langle z^2\rangle$ is the average value of $z^2$ calculated at equilibrium.  The term $\langle z^2\rangle$ ensures the orthogonality between the dipole compression mode and the center of mass oscillation. This is easily proven by noticing that the density variations $\partial_z [v n]$ associated with the DC mode give rise to a vanishing dipole moment: $\int dz z\partial_z [v n]=0$. 

Predictions \eqref{Eq:variationalLB} and \eqref{Eq:variationalDC} for the lowest breathing and the dipole compression modes are expected to provide an accurate approximation to the exact solutions of the hydrodynamic equation \eqref{Eq:wave equation} in all regimes of interaction and temperature. This is the consequence of the fact that the corresponding ansatz for the velocity field coincides with the exact solution of the hydrodynamic equation in important asymptotic regimes, where the equation of state exhibits a polytropic dependence on the density \cite{DeRosi2015}, like the $T=0$ weakly interacting limit, the $T=0$ Tonks-Girardeau limit  as well as in the classical regime of high temperatures \cite{highT, DeRosi2015}. One then expects that the same  ansatz for $v$ will be accurate also in the intermediate regimes of interaction and temperature. Such an accuracy was recently proven numerically by Hu \textit{et al.} \cite{Hu2014}. 
The values of the   hydrodynamic frequencies  calculated in the above three asymptotic regimes \cite{DeRosi2015} are reported  in  Table \ref{Tab:monopoledeep} for the lowest breathing mode  and in Table \ref{Tab:monopole2}  for the dipole compressional mode. 
Finally, we notice that the LB HD frequencies of Table \ref{Tab:monopoledeep} were obtained also by Bouchoule \textit{et al.} \cite{Bouchoule2016} using scaling arguments starting from the HD equations.

\begin{table}[h]
\caption{Hydrodynamic VS collisionless frequencies of the lowest breathing mode ($LB$) for a 1D Bose gas.}
\label{Tab:monopoledeep}
\begin{tabular}{cccc}
 \hline\noalign{\smallskip}
 \hline\noalign{\smallskip}
& \multicolumn{2}{c}{\textbf{Hydrodynamic}} \\
&\textbf{T = 0} &\textbf{high T} & \textbf{Collisionless}\\
\noalign{\smallskip}\hline\noalign{\smallskip}
\textbf{1D weakly interact. (BG)} &   $\sqrt{3}\omega_z$  &   $2\omega_z$ & $2\omega_z$ \\
\noalign{\smallskip}\hline\noalign{\smallskip}
\textbf{1D Tonks-Girardeau} &   $2\omega_z$  &   $2\omega_z$ & $2\omega_z$   \\
\noalign{\smallskip}\hline 
\noalign{\smallskip}\hline 
\end{tabular}
 \end{table}

\begin{table}[hb]
\caption{Hydrodynamic VS collisionless frequencies of the dipole compressional mode ($DC$) for a 1D Bose gas. }
\label{Tab:monopole2}
\begin{tabular}{cccc}
 \hline\noalign{\smallskip}
 \hline\noalign{\smallskip}
& \multicolumn{2}{c}{\textbf{Hydrodynamic}} \\
&\textbf{T = 0} & \textbf{high T} & \textbf{Collisionless}\\
\noalign{\smallskip}\hline\noalign{\smallskip}
\textbf{1D weakly interact. (BG)} &   $\sqrt{6}\omega_z$  &   $\sqrt{7}\omega_z$ & $3\omega_z\hspace{0.1cm} \& \hspace{0.1cm}1\omega_z$    \\
\noalign{\smallskip}\hline\noalign{\smallskip}
\textbf{1D Tonks-Girardeau} &   $3\omega_z$  &   $\sqrt{7}\omega_z$   & $3\omega_z\hspace{0.1cm} \& \hspace{0.1cm}1\omega_z$ \\
\noalign{\smallskip}\hline 
\noalign{\smallskip}\hline 
\end{tabular}
 \end{table}

\section{Sum rules and collective oscillations}
\label{Sec:sumrules}

Sum rules represent a powerful tool to describe the collective behaviour exhibited by quantum many-body systems \cite{Lipparini1989, Stringari1996, Pitaevskii2016}. Their main merit is that, in many cases, they provide accurate predictions for the collective frequencies avoiding the full solution of the quantum many-body problem. Furthermore, being based on the algebra of commutators, they emphasize the symmetry properties of the problem and the role of conservation rules. In general sum rules provide compact expressions for the $p$-moments
\begin{equation}
\label{Eq:mp}
m_p(F) =\hbar\int_{-\infty}^{+\infty} (\hbar \omega)^pS_F(\omega)d\omega
\end{equation}
of the dynamic structure factor  
\begin{equation}
\label{Eq:SFomega}
S_F(\omega)=Q^{-1} \sum_{n,m = 1}^N e^{-\beta E_m}|\bra{m} F \ket{n}|^2 \delta(\hbar\omega-\hbar\omega_{nm})\ ,
\end{equation}
where $F=\sum_{k=1}^N f(z_k)$ is the relevant excitation operator, $Q = \sum_{m=1}^N \exp[-\beta E_m]$ is the partition function and $\omega_{nm} = (E_n - E_m)/\hbar$ are the Bohr transition frequencies, relative to the Hamiltonian, Eq. \eqref{Eq:H}.

An important sum rule, widely employed in many-body calculations, concerns the inverse-energy weighted moment $m_{-1}$ of the dynamic structure factor. 
This moment is directly related to the static response $\chi(F)$
defined in  terms of the fluctuation $\delta \langle F\rangle =\lambda \chi(F)$, induced by an external static perturbation of the form $H_{pert}= -\lambda F$ applied to the system, according to the relationship \cite{Pitaevskii2016} $\chi(F) = 2m_{-1}(F)$.

The $m_{-1}$ sum rule  can be combined with the energy weighted sum rule, 
which in general can be reduced in the form of  a double commutator involving the Hamiltonian $H$ and the excitation operator $F$, yielding the simple result
\begin{multline}
\label{Eq:m1F}
m_{1}(F) =  \frac{1}{2} \langle [F, [H, F]] \rangle =   \frac{\hbar^2}{2m} N \langle |\nabla_z f(z)|^2 \rangle \ ,
\end{multline}
to provide an estimate of the collective frequency through the ratio 
\begin{equation}
\label{Eq:1-1F}
\hbar^2 \omega^2_{1,-1}= \frac{m_1}{m_{-1}}\ .
\end{equation}

In the presence of harmonic trapping, the choice for the excitation operator depends on the nature of the collective mode. For the lowest breathing mode the natural choice is provided by the operator $F_{LB}=\sum_{k=1}^N(z^2_k-\langle z^2 \rangle)$ \cite{note v} which ensures   the condition $\langle F_{LB} \rangle =0$ at equilibrium. In this case the inverse energy weighted moment can be easily calculated since the static perturbation $-\lambda F_{LB}$  consists of a simple renormalization of the harmonic trapping frequency. One then obtains the following result \cite{Pitaevskii2016, Menotti2002}
\begin{equation}
\label{Eq:m-1LB}
m_{-1}(LB)= -\frac{N}{m} \frac{\partial \langle z^2\rangle}{\partial \omega_{z}^2} \ ,
\end{equation}
for the inverse energy weighted moment.
On the other hand,  the energy weighted moment \eqref{Eq:m1F}, relative to the same excitation operator,  yields the result
\begin{equation}
\label{Eq:m1LB}
m_1(LB)= \frac{2N\hbar^2}{m} \langle z^2 \rangle \ ,
\end{equation}
so that the ratio between the two sum rules provides the  expression 
\begin{equation}
\label{Eq:1-1LB}
\omega^2_{1,-1}(LB)= -2 \frac{\langle z^2\rangle}{\partial \langle z^2\rangle/\partial \omega_z^2} 
\end{equation}
for the squared collective frequency.

Result \eqref{Eq:1-1LB} was successfully employed to evaluate the LB frequency in 1D Bose gases at zero temperature \cite{Menotti2002}. In particular, by using the Local Density Approximation to evaluate the $\omega_z$-dependence of the average square radius, this equation accounts for the transition of the collective frequency from the value $\sqrt{3}\omega_z$ holding in the weakly interacting Bose gas to the value $2\omega_z$ holding in the Tonks-Girardeau limit, see Table \ref{Tab:monopoledeep}. Since Eq. \eqref{Eq:1-1LB} does not assume the Local Density Approximation, it can be also used to estimate the collective frequencies when the coupling constant $g_{1D}$ or the number of atoms are small \cite{Gudyma2015, Gudyma20152}. One should however notice that result  \eqref{Eq:1-1LB} is not adequate to describe
the frequency of the LB mode at finite temperature. This is best understood in the classical  limit of high temperatures where  Eq. \eqref{Eq:1-1LB} provides the result $\sqrt 2 \omega_z$ for the collective frequency to be compared with the exact value $2 \omega_z$ holding in the classical limit where the Hamiltonian of the system reduces to the ideal gas value (see Table \ref{Tab:monopoledeep}). The discrepancy between the two values is 
due to the fact that, at finite temperature, the operator $F_{LB}$ excites zero frequency modes which provide a finite contribution to the inverse energy weighted moment sum rule \cite{uniform}.    

The correct value of the collective frequency at finite temperature is recovered if, instead of calculating the inverse energy weighted sum rule, one evaluates the cubic energy weighted sum rule $m_3(F)$ which can be written in the form of a double commutator involving the Hamiltonian $H$ and the commutator $[H,F]$: 
\begin{equation}
\label{Eq:m3F}
m_{3}(F)  = \frac{1}{2}\langle[[F,H],[H,[H,F]]]\rangle \ .
\end{equation}
Differently from $m_{-1}(F)$, the cubic energy weighted moment is not sensitive to the zero frequency modes excited by the operator $F$ at high temperature. Evaluation of the triple commutator \eqref{Eq:m3F} with the Lieb-Liniger Hamiltonian \eqref{Eq:H}  yields the following result for  
the $m_3$ sum rule relative to the excitation operator $F_{LB}=\sum_{k = 1}^N(z^2_k-\langle z^2 \rangle)$:
\begin{equation}
\label{Eq:m3LB}
m_{3}(LB) =\frac{2\hbar^4}{m^2}\left(4\langle H_{kin}\rangle + 4\langle H_{trap} \rangle + \langle H_{int} \rangle \right)\ .
\end{equation}

A useful simplification of Eq. \eqref{Eq:m3LB}  is provided by the virial theorem \cite{Pitaevskii2016, Stringari1996, Gudyma20152}, which can be derived by imposing  the general condition 
$\langle [H,G] \rangle =0$ holding at  equilibrium for any choice of the operator $G$. By making the choice $G= \sum_{k = 1}^N \left(z_kp_{z,k} +p_{z,k}z_k\right)$ corresponding to a scaling deformation of the many-body wave function, one derives the exact relationship 
\begin{equation}
\label{Eq:virial}
2\langle H_{kin} \rangle - 2 \langle H_{trap} \rangle + \langle H_{int} \rangle=0\ .
\end{equation}

Thanks to the virial theorem \eqref{Eq:virial} the cubic energy weighted sum rule \eqref{Eq:m3LB}
can be further simplified and, combined with the energy weighted sum rule \eqref{Eq:m1LB}, yields the following expression for the LB collective frequency \cite{Gudyma20152}
\begin{equation}
\label{Eq:31LB}
\hbar^2\omega_{3,1}^2(LB)= \frac{m_{3}(LB)}{m_1(LB)} = \hbar^2 \omega_z^2 \left( 4 - \frac{\langle H_{int}\rangle}{2\langle H_{trap} \rangle} \right)\ ,
\end{equation}
or, equivalently \cite{Gudyma20152},
\begin{equation}
\label{Eq:31LBbis}
\omega_{3,1}^2(LB) =  \omega_z^2\left( 3 + \frac{\langle H_{kin}\rangle}{\langle H_{trap} \rangle} \right)\ ,
\end{equation}
holding also beyond LDA.
Eq. \eqref{Eq:31LB}  explicitly shows that, if the average value of the  interaction energy is  negligible, as happens in the TG regime and in  the    collisionless regime of high temperatures, one recovers the correct value $2\omega_z$ for  the lowest compression mode (see Table \ref{Tab:monopoledeep}). In the case of the weakly interacting Bose gas one can neglect, at T=0, the kinetic energy term and Eq. \eqref{Eq:31LBbis} correctly reproduces the hydrodynamic value $\sqrt{3}\omega_z$.
In conclusion one expects that the sum rule result $m_3/m_1$  will provide an excellent estimate of the frequency of the lowest compression mode in all ranges of temperature, interaction and number of particles. At $T=0$ it is expected to provide results of similar accuracy as prediction \eqref{Eq:1-1LB} based on the ratio between the energy weighted and inverse energy weighted sum rule.  The expression \eqref{Eq:31LB} for the LB collective frequency was already considered by Fang \textit{et al.} \cite{Fang2014} to analyze their experimental data at finite temperature.

A further interesting expression for the $\omega_{3,1}^2$ ratio can be obtained by using the Hellmann-Feynman expression $\langle H_{int}\rangle = g_{1D}\partial F/\partial g_{1D}$ for the interaction energy, where $F$ is the free energy of the system. In this way Eq. \eqref{Eq:31LB} takes the form

\begin{equation}
\label{Eq:Referee2}
\omega_{3,1}^2(LB) = \omega_z^2 \left[4 + \frac{\hbar^2 \mathcal{C} a_{1D}}{2m \langle H_{trap} \rangle}   \right]
\end{equation}

where we have introduced the 1D Tan's contact parameter $\mathcal{C}= (m/\hbar^2) \partial F/\partial a_{1D}$ with  $a_{1D}=-2\hbar^2/mg_{1D}$ the 1D scattering length. The same result can be obtained by using the Tan's contact 1D virial theorem (see, for example \cite{Valiente2012}). The  Tan's contact,  which characterizes  the large momentum tail of the momentum distribution, can be also expressed in terms of the pair correlation function \cite{Gangardt2003}. Result \eqref{Eq:Referee2} relates the frequency of the lowest compression modes, fixed with high accuracy by the ratio $\omega_{3,1}^2(LB)$, to independently measurable quantities.

A similar analysis can be worked out for the dipole compression mode excited by the operator $F_{DC}=\sum_{k = 1}^N f_{DC}(z_k)$ with $f_{DC}(z)=z^3/3 -z\langle z^2 \rangle$ \cite{note v}. The choice ensures that the  operator $F_{DC}$ will not excite the center of mass (dipole) oscillation. This can be easily shown by checking that the crossed energy weighted sum rule $\langle [F_{D},[H,F_{DC}]]\rangle$, with $F_{D}=\sum_{k = 1}^N z_k$, identically vanishes. 

In the case of the DC mode the static response, and hence the inverse energy weighted sum rule, can be easily calculated only in the LDA where, in the presence of the external perturbation $-\lambda F_{DC}$, the chemical potential is modified according to $\mu \to \mu -\lambda f_{DC}(z)$ and the density profile is, accordingly, modified as $ n(z) \to n(z) +\lambda f_{DC}(z) (\partial n/\partial \mu)_T$. The inverse energy weighted sum rule relative to the DC mode then takes the useful form \cite{notedeltamu}:
\begin{equation}
\label{Eq:m-1 general}
m_{-1}(DC) = \frac{1}{2}   \int dz \left(\frac{z^3}{3} - z \langle z^2 \rangle   \right)^2\left(\frac{\partial n}{\partial \mu}\right)_T \ .
\end{equation}
Using Eq. \eqref{Eq:m1F}, the energy weighted moment is also easily evaluated and takes the form: 
\begin{equation}
\label{Eq:m1 DC}
m_1(DC) = \frac{\hbar^2N}{2m}(\langle z^4\rangle - \langle z^2\rangle^2) \ .
\end{equation}

It is straightforward to verify that, at T=0, the ratio $m_1/m_{-1}$ provides the correct (squared) hydrodynamic frequencies both in the weakly interacting Bose gas ($\sqrt{6}\omega_z$), where $\partial \mu/ \partial n = g_{1D}$, and in the Tonks-Girardeau limit ($3\omega_z$), where $\partial \mu/ \partial n = \hbar^2 \pi^2 n/m$. At high temperatures, where $\partial \mu/ \partial n_{|T} = \partial P/ \partial n_{|T}/n = k_BT/n$,
one instead finds that the frequency $\omega_{1,-1}$  takes the value $\sqrt{3}\omega_z$ which is smaller than the hydrodynamic value $\sqrt{7}\omega_z$,  similarly to the case of the LB mode discussed above. This result is the consequence of the fact  that the DC operator $F_{DC}$ excites, at high temperature, two modes with frequency equal to $\omega_z$ and $3\omega_z$, respectively. The corresponding strengths $\sigma_1$ and $\sigma_3$ characterizing the dynamic structure factor $S_F(\omega)$  can be easily evaluated through the identification 
\begin{equation}
\label{Eq:1-1DC}
\omega_{1,-1}^2(DC)= \frac{\sigma_1\omega_z+3\sigma_3\omega_z}{\sigma_1/\omega_z+\sigma_3/3\omega_z}= 3 \omega^2_z\ ,
\end{equation}
yielding the relationship $\sigma_1=\sigma_3$.
The above result for the strengths $\sigma_1$ and $\sigma_3$ permits to predict, in the same regime of high temperature,  the value of the ratio between the cubic and the energy weighted moments. We find
\begin{equation}
\label{Eq:31DC}
\omega_{3,1}^2(DC)= \frac{\sigma_1\omega_z^3+27\sigma_3\omega_z^3}{\sigma_1\omega_z+3\sigma_3\omega_z}=7\omega_z^2 \, .
\end{equation}

As in the case of the LB mode also for dipole compression mode the cubic energy weighted sum rule   can be   calculated on a general basis in all regimes of temperature by carrying out explicitly the algebra of commutators. We find the result
\begin{multline}
\label{Eq:general m3 DC}
m_3(DC) = \frac{\hbar^4N}{m^2}[g_{1D}\langle z^2 \rangle \langle \delta(z_{ij})\rangle + g_{1D}\langle Z_{ij}^2 \delta(z_{ij}) \rangle - \\ - \frac{3}{2}m\omega_z^2 \langle z^2 \rangle^2 +  \frac{1}{m}\langle z^2 \rangle \langle p_z^2 \rangle + \frac{3}{m}\langle p_z z^2 p_z \rangle + \frac{3}{2}m\omega_z^2 \langle z^4\rangle - \frac{\hbar^2}{m}]\ ,
\end{multline}
where $Z_{ij}=(z_i+z_j)/2$ is the center-of-mass coordinate and we have defined the intensive quantities $\langle \delta(z_{ij})\rangle  \equiv \langle \sum_{i>j}^N\delta(z_{ij})\rangle /N$ and $\langle Z^2_{ij}\delta(z_{ij})\rangle  \equiv \langle\sum_{i>j}^N Z^2_{ij}\delta(z_{ij})\rangle /N$.
Similarly to the case of the LB mode discussed above, also for the DC mode one can obtain a useful relationship among the various  contributions entering \eqref{Eq:general m3 DC} with the help of a generalized virial theorem derivable by imposing the condition
$\langle [H,G] \rangle =0$, with the choice $G= \sum_{k = 1}^N \left(z_k^3p_{z,k} +p_{z,k}z_k^3\right)$. This yields the relationship: 
\begin{equation}
\label{Eq:supervirial}
\frac{6}{m}\langle p_z z^2 p_z \rangle + 6g_{1D} \langle \delta(z_{ij})Z_{ij}^2\rangle -2m\omega_z^2 \langle z^4 \rangle - \frac{3\hbar^2}{m} = 0\ .
\end{equation}
It is easy to verify that the ratio $m_3/m_1$ provides the correct square excitation energy in some relevant limits at zero temperature. These include the weakly interacting Bogoliubov gas, where the kinetic energy contribution to  \eqref{Eq:virial}, \eqref{Eq:general m3 DC} and \eqref{Eq:supervirial} vanishes and the DC excitation  frequency takes the T=0 hydrodynamic  value $\sqrt{6}\omega_z$, and in the Tonks-Girardeau limit, where the contribution due to the interaction   vanishes and the frequency takes the value $3\omega_z$ \cite{Hu2014, DeRosi2015}. At T=0 the ratio $m_3/m_1$ also accounts for the regimes of small coupling constant $g_{1D}$ or small  atomic numbers $N$ where the LDA is no longer applicable \cite{Chen2015}.  
At high temperature, where interaction effects are negligible, the ratio $m_3/m_1$ reproduces the hydrodynamic result  $\sqrt{7}\omega_z$  for the average excitation frequency, consistently with the derivation of result \eqref{Eq:31DC}.

In the next Section we will provide a more detailed description of the excitation spectrum of the dipole compression mode, by studying the response of the trapped gas to a sudden density perturbation, giving rise to observable signatures of the collisional VS collisionless nature of the gas.

\section{Exciting the dipole compression mode}
\label{Sec:Compressiondipole}

\begin{figure}[htbp]
\centering
\includegraphics[scale=0.5]{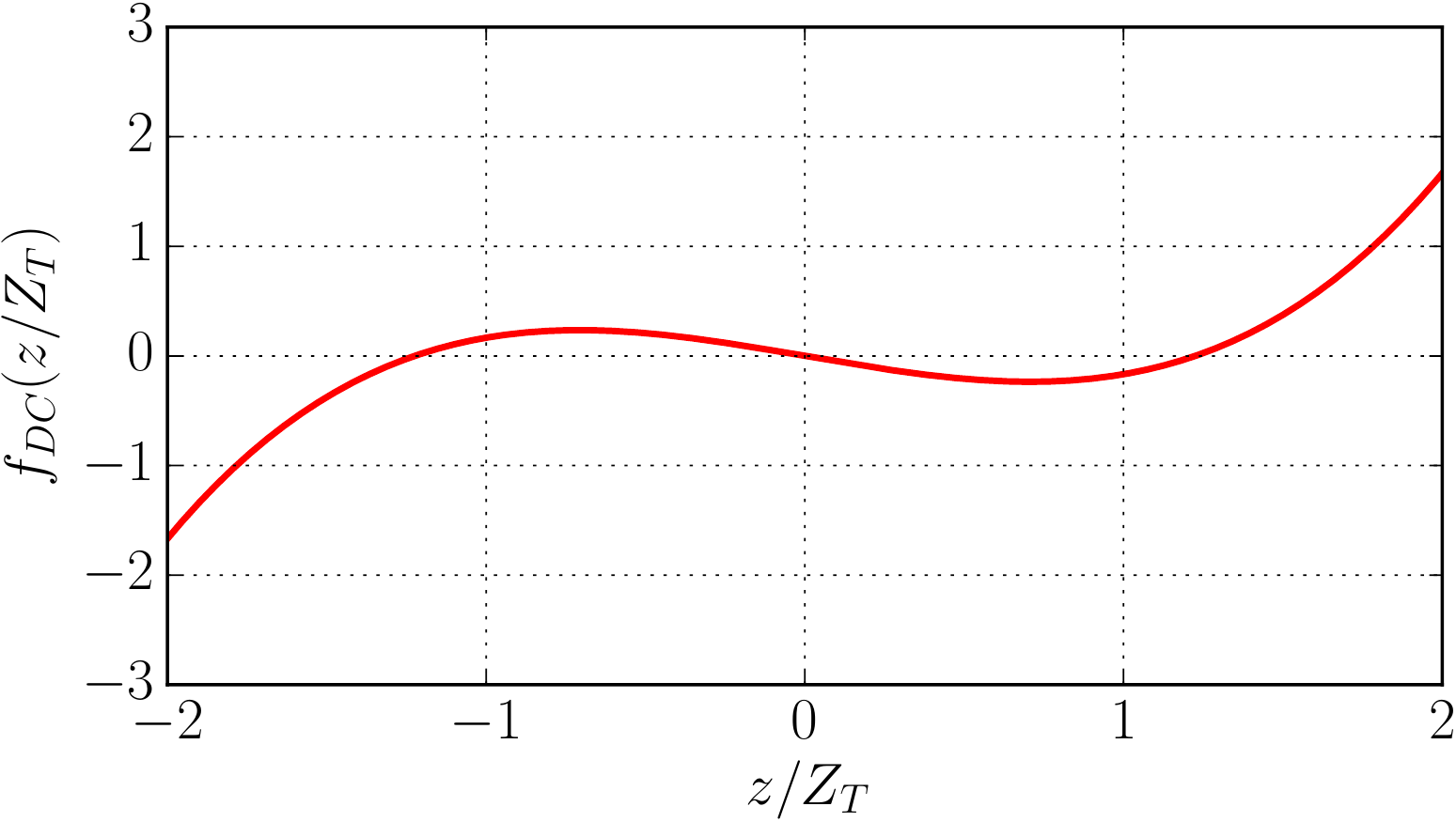}
\caption{External perturbation $f_{DC}(z)= z^3/3-z\langle z^2\rangle$ exciting the dipole compression (DC) mode. The value of $\langle z^2\rangle$ was calculated using a Maxwell--Boltzmann distribution with thermal radius $Z_T = \sqrt{2k_BT/(m\omega_z^2)}$. 
 \label{fig:fDC}}
\end{figure}

In this Section we exploit the peculiar behaviour exhibited by the dipole compression mode resulting from a sudden small density perturbation of the form $H_{pert}(z,t) = \lambda F_{DC}(z) \Theta(t)$ with $F_{DC} = \sum_{k = 1}^N f_{DC}(z_k)$, $f_{DC}(z)= z^3/3-z\langle z^2\rangle$ and $\Theta(t)$ the Heaviside function. Perturbations of similar form can be tailored with laser techniques and have been already implemented in the case of highly elongated Fermi gases \cite{Tey2013}. 
The form of the DC perturbation $f_{DC}(z)$ is shown in Fig. \ref{fig:fDC} where we have expressed the variable $z$ in units of the thermal radius $Z_T = \sqrt{2k_BT/(m\omega_z^2)}$. As pointed out in the previous Section, the excitations  produced by this perturbation are exactly decoupled from the center of mass motion.

\begin{figure}[htbp]
\centering
\includegraphics[scale=0.5]{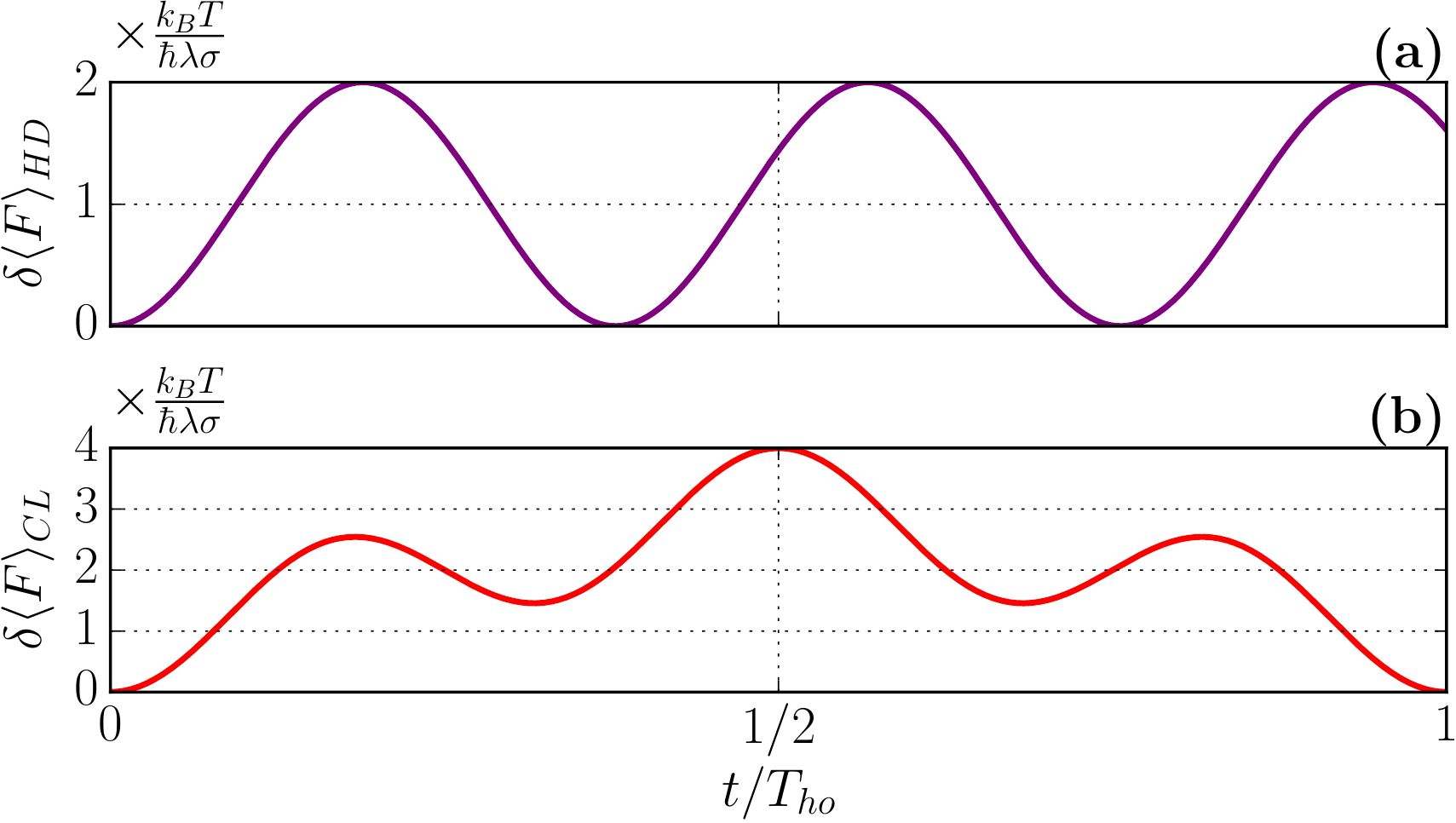}
\caption{Time evolution of the expectation value $\delta \langle F \rangle$, in units of oscillator time $T_{ho} = 2\pi/\omega_z$, following the perturbation of the dipole compression mode (see text).  \\ In  the hydrodynamic regime of high temperatures \textbf{(a)} the signal is characterized by the single frequency $\sqrt{7}\omega_z$, while in the collisionless regime of high T \textbf{(b)} by a periodic beating of the 2 frequencies $\omega_z$ and $3\omega_z$.
 \label{fig:HD_CL}}
\end{figure}

According to linear response theory \cite{Pitaevskii2016} the time evolution of the expectation value $\delta \langle F \rangle(t) = \int dz \delta n(z, t)f_{DC}(z)$ follows the law \cite{Zambelli2001}
\begin{equation}
\label{Eq:ft)}
\delta \langle F \rangle(t) = \frac{\lambda \hbar}{k_BT} \int_{-\infty}^{+\infty} d\omega' S_F(\omega') \left[1 - \cos(\omega't) \right] \ ,
\end{equation}
where $S_F(\omega)$ is the dynamic structure factor relative to the excitation operator $F$, see Eq. \eqref{Eq:SFomega}. In the hydrodynamic regime a single frequency, provided by Eq. \eqref{Eq:variationalDC}, will appear in the time evolution of the signal. According to the results of Table \ref{Tab:monopole2}, this frequency will evolve continuously from the low temperature T value $\sqrt 6 \omega_z$ (weakly interacting limit) or $3 \omega_z$ (Tonks-Girardeau limit) to the large T value $\sqrt 7\omega_z$. In Fig. \ref{fig:HD_CL}\textbf{(a)} we show the time dependence of the signal $\delta \langle F \rangle(t)$ predicted in the high $T$ hydrodynamic limit. If instead the system is in the collisionless regime of high temperature, and hence $S_F(\omega) = \sigma [\delta (\omega -\omega_z) +\delta(\omega- 3 \omega_z) + \omega \to - \omega]$ (we have set $ \sigma_1 = \sigma_3 \equiv \sigma$, according to  the discussions presented at the end of the previous Sec. \ref{Sec:sumrules}), the signal will exhibit a typical beating involving the two frequencies, as reported in \ref{fig:HD_CL}\textbf{(b)}.

The observation of the transition between a single frequency signal to the beating regime can then be considered a signature of the transition between the hydrodynamic to the collisionless regime. A transition of similar nature was observed   in the study of the scissors mode of 3D Bose gases in  a deformed harmonic potential where the frequency has a single value at low temperature in the superfluid Bose-Einstein condensed phase, while the spectrum exhibits a beating between two frequencies for temperatures  larger than the critical temperature where the system is in the non superfluid collisionless regime \cite{GueryOdelin1999, Marago2000}.

\section{Conclusions}
\label{Sec:conclusion}
In this paper we have calculated the collective frequencies of a 1D harmonically trapped Bose gas in different regimes of interaction, temperature and number of particles.

We have developed two different theoretical methods: 
 the hydrodynamic approach, rewritten in an easier variational formulation, and the more microscopic  sum-rule approach. While the first method can be applied only within  the Local Density Approximation (LDA) and enables us to calculate the hydrodynamic frequencies for all interaction and temperature regimes, the sum-rule approach allows us to calculate  the collective frequencies even beyond the  LDA and in the collisionless regime of high temperatures.  

The inverse energy weighted ($m_{-1}$), the energy weighted ($m_{1}$) and the cubic energy weighted  ($m_3$) sum rules are calculated and their applicability to exploit the behaviour of the collective frequencies at zero as well as at finite temperature have been explicitly discussed. We have furthermore developed the formalism of the virial theorem which permits to derive  more compact expressions for the average excitation frequencies, defined through the ratio  $\hbar^2 \omega^2 = m_3/m_1$.

The combined use of the hydrodynamic and sum rule approaches enables us to draw important conclusions about the temperature dependence of the collective frequencies. While in the case of the lowest breathing mode the frequencies in the high temperature hydrodynamic and collisionless regimes coincide and are equal to $2\omega_z$, where $\omega_z$ is the oscillator frequency, a different scenario emerges in the case of the dipole compression mode excited by the operator $f_{DC}(z) = z^3/3-z\langle z^2\rangle $. In the dipole compression case, the hydrodynamic approach in fact predicts the value $\sqrt{7}\omega_z$ for the collective frequency, while in the collisionless regime the same operator gives rise to the excitation of two different frequencies given by $\omega_z$ and $3\omega_z$.  By calculating the response of the system to a sudden perturbation of the form $ \lambda f_{DC}(z) \Theta (t)$, we predict a typical beating between the two frequencies whose experimental observation would provide a useful signature of the achievement of the collisionless regime. 
The investigation of the temperature dependence of the dipole compression mode is then expected to provide valuable information on the transition between the hydrodynamic and collisionless regime  and  on the role of collisions in 1D interacting Bose gases.

The sum rule approach is also expected to provide a useful tool to explore the behaviour of the dipole compression frequencies when the Local Density Approximation is not available at zero as well as at finite temperature and for different interaction regimes. This will be the object of a future investigation.

\begin{acknowledgments}
Fruitful discussions with C. Salomon, S. Giorgini and C. Menotti are acknowledged. The Authors are grateful to the Referee of this paper for suggesting the inclusion of Eq. \eqref{Eq:Referee2} in the text. This work has been supported by ERC through the QGBE grant, by the QUIC grant of the Horizon2020 FET program and by Provincia Autonoma di Trento.
\end{acknowledgments}


\begin{thebibliography}{99}
\bibitem{highT}
High temperature regime implies that the thermal energy is much higher than the degeneracy energy $E_{deg} \sim \hbar^2 n^2/m$: $k_BT \gg E_{deg}$. On the other hand, the temperature should not be too high in order to ensure the 1D condition: $k_BT \ll \hbar \omega_\perp$, where $\omega_\perp$ is the radial trapping frequency.
\bibitem{note v}
The excitation operator $F = \sum_{k = 1}^N f(z_k)$ is related to the velocity field $v$, defined in Sec. \ref{Sec:variational}, by $v(z) \propto \nabla_z f(z)$.
\bibitem{notedeltamu}
The same procedure, applied to the lowest breathing mode, should take into account a further position independent correction $\delta \mu = - \int dz f(z) (\partial n/\partial \mu)_T/ \int dz (\partial n/\partial \mu)_T$,  which is required to ensure the particle number conservation $\int dz \delta n = 0$. This yields  the general expression 
\begin{equation}
\label{Eq:m-1 generalA}  
m_{-1} = \frac{1}{2}   \int dz f(z)\left[f(z) + \delta \mu\right]\left(\frac{\partial n}{\partial \mu}\right)_T\ ,
\end{equation}
for the inverse energy weighted sum rule holding, in the LDA, for any choice of $f(z)$.
\end{thebibliography}
\end{document}